\documentclass[pdflatex, sn-mathphys-num, ]{sn-jnl}

\usepackage{graphicx}%
\usepackage{multirow}%
\usepackage{amsmath,amssymb,amsfonts}%
\usepackage{amsthm}%
\usepackage{mathrsfs}%
\usepackage[title]{appendix}%
\usepackage{xcolor}%
\usepackage{textcomp}%
\usepackage{manyfoot}%
\usepackage{booktabs}%
\usepackage{algorithm}%
\usepackage{algorithmicx}%
\usepackage{algpseudocode}%
\usepackage{listings}%
\usepackage{caption}
\usepackage{subcaption}
\usepackage{tabularx}
\usepackage{booktabs}
\usepackage{tabularx}
\usepackage{booktabs}
\usepackage{multirow}
\usepackage{float}

\usepackage[dvipsnames]{xcolor}

\begin{document}

\title[Article Title]{On the relative CNO underabundance in quasar absorption systems at $z \sim 3$ arising from Population III enrichment and attenuation by intermediate-mass black holes and primordial baryon accretion}


\author*[1]{\fnm{Murilo} \sur{Macedo}}\email{murilo.macedo@inpe.br}

\author[1]{\fnm{Carlos Alexandre} \sur{Wuensche }}\email{ca.wuensche@inpe.br}
\equalcont{These authors contributed equally to this work.}

\author[1]{\fnm{Oswaldo Duarte } \sur{Miranda }}\email{oswaldo.miranda@inpe.br}
\equalcont{These authors contributed equally to this work.}

\affil*[1]{\orgdiv{National Institute for Space Research (INPE)}, 
\orgname{Instituto Nacional de Pesquisas Espaciais}, 
\orgaddress{\street{Av. dos Astronautas, 1758}, \city{São José dos Campos}, \postcode{12227-010}, \state{São Paulo}, \country{Brazil}}}


\abstract{% o que fazemos:
This article uses an adapted version of the semi-analytical model of cosmic chemical enrichment developed by \citet{Corazza_2022} to reproduce the observed abundances of C, N, and O in absorption systems of quasar spectra (ASQS) at $z \gtrsim 3-6$, addressing an overproduction issue of the abovementioned elements.
% Como fazemos:
 We address this discrepancy by updating the cosmic star formation rate (CSFR) and introducing intermediate-mass black holes (IMBHs) as permanent matter sinks 
 without accounting for a dynamic cosmic mass accretion rate. 
 Our results indicate that IMBHs act as essential metallicity attenuators 
through mass sequestration, providing the physical regulation necessary to reconcile theoretical yields with observed data. We show that the interplay between Pop III yields, the cosmic baryon accretion rate (CBAR) from primordial nucleosynthesis, and mass sequestration by IMBHs mitigates the CNO excess.
 % o que concluímos:
 This work reinforces the role of black hole-driven processes in the chemical evolution of the Universe and identifies IMBH accretion rates as a primary area for future refinement.
}


\maketitle

\section{Introduction}
\label{sec:intro}

According to the Standard Cosmological Model, in the first few minutes after the Big Bang, the universe had sufficiently high temperatures for the formation of atomic nuclei through nuclear fusion, resulting in primordial chemical abundances of approximately 25\% Helium ($^4He$) and 75\% hydrogen ($H$)  \citep{Particle_review_2022, Kurichin_2022, Kurichin_2021, Dodelson_2021, Planck2018_Parameters, Weinberg_2008}. As the universe expands, dark matter clusters into halos, around which primordial gas is continuously attracted. The cosmic ``baryon accretion rate'' (hereafter BAR), which describes how this gas is incorporated into dark matter halos, serves as fuel for star formation and as a moderator of the metallicity of the medium. Primordial BAR constitutes the initial step in cosmic chemical enrichment.

\citet{Corazza_2022} investigate the potential contribution of intermediate-mass Population III and II stars to cosmic chemical enrichment at $z \lesssim 4$ using absorption systems in quasar spectra (hereafter ASQS) as the primary observational constraint \citep{Wolfe_2005}. The most significant discrepancies between their enrichment modeling and the observations is the theoretical overproduction of carbon (C), nitrogen (N), and oxygen (O), coming from five  different initial mass function (hereafter IMF). Even when accounting exclusively for Population III (Pop III) stellar yields, these models still produce CNO excess. This persistent excess suggests that relying solely on primordial stellar nucleosynthesis may be insufficient to explain the chemical signatures of these systems. 

The chemical evolution of CNO for some metal-poor ASQS is frequently modeled through the %enrichment contributions of the first stars 
contributions of the first stars to the cosmic enrichment \citep{Sodini_A_2024, Salvadori_2023, Welsh_L_2022, Cooke_2011_the_most_metal, Cooke_2011_b, Kobayashi_2011, Pettini_2008}. Building on the premise that Pop III stars may play a significant role in the enrichment of some $z \gtrsim 2$ absorption systems, this work proposes a solution to the persistent CNO overabundance identified in the models of \citet{Corazza_2022}.

We address this discrepancy by introducing two fundamental refinements to the chemical evolution framework: first, an updated cosmic star formation rate; and second, the inclusion of the mass fraction consumed by intermediate-mass black holes (hereafter IMBHs) with masses up to $10^{4}\,M_{\odot}$ taken as an upper mass limit and their probability of occurrence described by the IMF \citep{Abbott_2022, Greene_2020, Pereira_Miranda_2011}. 
In this framework, IMBHs are treated as permanent sinks of mass. By assuming that stars in this mass range produce no chemical yields and remain entirely as black hole remnants, we focus on the mass-locking effect as the primary regulatory mechanism for cosmic metallicity.
Rather than treating the excess CNO as an inherent feature of primordial nucleosynthesis, our results suggest that the inclusion of black hole-driven processes provides the significant physical regulation to reconcile theoretical yields with the observed data even for more recent identified systems at $z\sim6$. %Em particular, os resultados demonstram consistencia com o trabalho de \citet{Sodini_A_2024}.

Thus, in this work we propose two processes that attenuate the metallicity of the medium: black holes and the  CBAR. The role of black holes as metallicity attenuators has likewise been identified as necessary in earlier work \cite{Isobe_2023}. More recently, \citet{Zhu_2025} associated the presence of black holes with the suppression of the star formation rate which we interpret as leading to a consequent decrease in stellar nucleosynthesis and chemical enrichment. In this specific framework, IMBHs are treated primarily as matter sinks, focusing on the impact of mass and metal removal from the interstellar medium rather than the complexities of radiative feedback. 

After this introduction, we present a brief discussion of the CBAR and the cosmic star formation rate (hereafter CSFR) and the functional form that best fits the observational data up to $z \sim 15$ in section 
\ref{sec:csfr}.

Section \ref{sec:themodel} describes the semi-analytical model of cosmic chemical enrichment is briefly introduced, in which we highlight the fundamental nucleosynthetic contribution provided by the inclusion of Pop III stars in our model.

In section \ref{sec:results} we  compare the results of our model for the prediction of CNO abundance with measurements from quasar absorption-line systems \citep{Houjun_Mo_2010, Khare_2007, Wolfe_2005}. 

Finally, in section \ref{sec:conclusions} we provide a brief summary of our results, which demonstrate that IMBHs and Pop III stars likely play a fundamental role in the chemical enrichment of ASQS, particularly those at high redshifts ($z \gtrsim 6$).

%\textcolor{blue}{\citet{Suarez-Andres_2017_C, Suarez-Andres_2016_N} investigaram a o comportamento da abundância de C, N e O em estrelas hospedeiras de planetas e concluíram que elas tendem a ser mais enriquecidas, daquelas que não possuem planetas. Corpos menores poderiam funcionar como sumidouros de matéria, em particular, CNO.}

%\alex{\citet{Corazza_2022} investigate the potential contribution of intermediate-mass Population III and II stars to cosmic chemical enrichment at $z \lesssim 4$ using absorption systems in quasar spectra (hereafter ASQS) as the primary observational constraint \citep{Wolfe_2005}. The most significant discrepancies between their enrichment modelling and the observations is the theoretical overproduction of carbon (C), nitrogen (N), and oxygen (O), coming from five  different initial mass function (hereafter IMF).

%Highly carbon-enriched planets have also been investigated in previous studies \citep{Madhusudhan_2012, Madhusudhan_2011, Hebb_2009}. However, the planet hypothesis can be ruled out in light of the low estimates of the cosmic planet formation rate \citep{Lapi_2024, Behroozi_2015}, as planets would represent a very small mass fraction, insufficient to alleviate the discrepancy in the Corazza model.

\section{Primordial baryon and cosmic star formation rates}
\label{sec:csfr}

%------------------------------------
% BAR

Primordial gas is continuously incorporated into dark matter halos, thus serving as a continuous source of fuel for star formation. The amount of primordial matter per unit mass and time that is added to the halos is referred to as the CBAR is denoted by $a_b(z)$ \citep{Daigne_2006}. This rate is closely linked to models of large-scale structure formation in the Universe, as it depends on the shape and number in which the halos are distributed throughout the Universe. Based on the Press–Schechter formalism \citep{Press_Schechter_1974}, \citet{Pereira_Miranda_2010} derived a cosmic BAR, which has subsequently been employed in several studies \citep{Miranda_2025, Corazza_2022, Gribel_2017, Pereira_Miranda_2015, Pereira_Miranda_2011} and its redshift dependence is given by
\begin{equation}
a_{b}(z)=\frac{0.0494(1+z)^{1.395}}{1+\left[\left(1+z\right)/16.642\right]^{9.468}}\; [M_{\odot} \text{ year}^{-1} \text{ Mpc}^{-3}].
\label{BAR_equation}
\end{equation}
We adopt this same BAR formulation in the present work, ensuring consistency with previous studies.
%------------------------------------
% CSFR
The CSFR quantifies the amount of gas and dust mass that transforms into stars per unit time and volume in the Universe \citep{Corazza_2022, Schneider_2016, Madau_2014}. This rate is fundamental for understanding cosmic chemical enrichment and the evolution of galaxies, while CBAR provides fuel from primordial nucleosynthesis for star formation. 

When observing the spectrum of a galaxy, which consists of the superposition of the spectrum of each star within that galaxy, it is expected that, over time, there will be a shift in the dominance of various stellar populations. In the early epoc of the galactic evolution one should expect to measure ultraviolet (UV) emission from young stars; an increase in emission in the infrared (IR) range should dominate as red giants emerge; and after $\sim 3 \times 10^{9}$ years, an increase in UV emission due to hotter stars on the horizontal branch and the emergence of white dwarfs \citep{Robertson_2022, Schneider_2016, Madau_2014} are the main contributors to the galactic spectrum. In other words, one way to estimate the CSFR is by inferring stellar mass from the measured luminosity density of various galaxies at different redshifts (the galaxy luminosity function), where this inference can be made in the light of stellar population models \citep{Robertson_2024, Schneider_2016, Madau_2014}. A more detailed description of these models can be found in \cite{Johnson_2021, Eldridge_2017, Conroy_2009}.

\begin{figure}[h]
    \centering
    \includegraphics[width=0.9\linewidth]{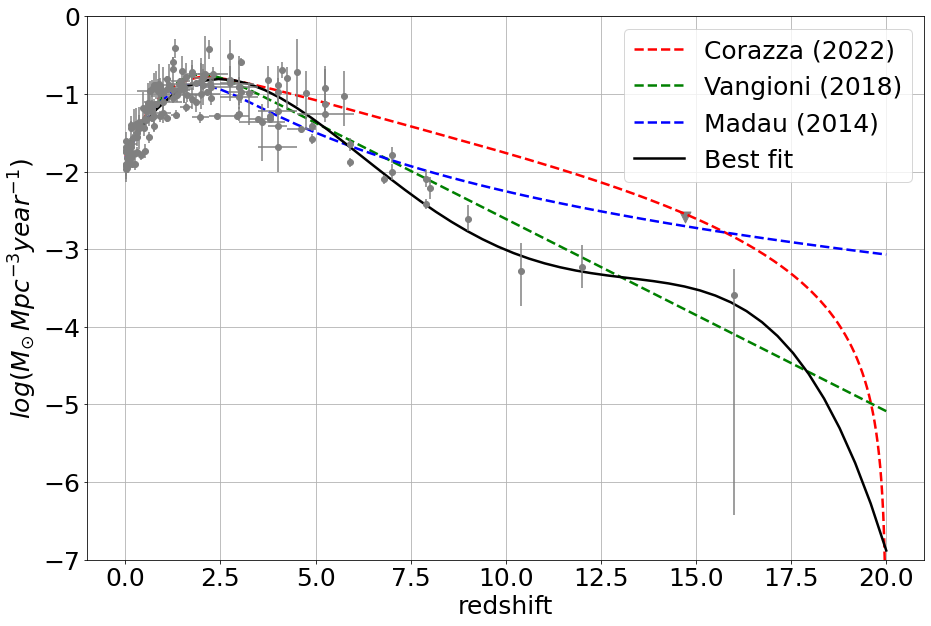}
    \caption{CSFR from \citet{Corazza_2022}, \citet{Vangioni_2018}, and \citet{Madau_2014}. The Madau \& Dickinson CSFR has been extrapolated to $z>6$. %, and the open triangle at $z \sim 15$ denotes an upper limit. 
    The black line denotes the best fit to the observational data (gray points) from \citep{Bouwens_2023, Harikane_2023, Katsianis_2021, Gruppioni_2020, Bouwens_2020, Driver_2018, Rowan-Robinson_2016, Madau_2014, Hopkins_2004}. The open triangle at $z \sim 15$ denotes an upper limit.}
    \label{fig_csfr}
\end{figure}

Pop III stars are expected to have formed in dark matter halos from catastrophic collapses of primordial gas, starting at redshift $z \sim 20$; we conservatively assume, following the results in the literature, that Pop III star formation ends at $z=6$ \citep{Maiolino_2024, Chantavat_2023, Donnan_2022, Corazza_2022}. The model used in this article assumes that primordial stars began to form at $z=20$, which is consistent with the model for the CBAR from \cite{Pereira_Miranda_2010}. In this context, the second fundamental element in building an enrichment model is the CSFR, denoted by $\dot{\rho}_{\star}(t)$. This quantity measures the amount of mass per unit time and volume that is converted into stars \citep{Schneider_2016, Madau_2014, Matteucci_2012}. Figure \ref{fig_csfr} illustrates the best-fit curve to the observational data used in this work. The best-fit curve to the observational data, expressed in $\log(M_{\odot}\,\text{yr}^{-1}\,\text{Mpc}^{-3})$, used in this work, is given by the following polynomial, where $z$ represents the redshift
\begin{equation}
log_{10}(\dot{\rho}_{\star}(z))=-4.15\times10^{-4}z^{4}+1.70\times10^{-2}z^{3}-2.25\times10^{-1}z^{2}+8.36\times10^{-1}z-1.75.
\label{melhor ajuste}
\end{equation}
The total number of stars and their mass distribution dictate the rate at which chemical elements are processed in stellar interiors and subsequently ejected into the interstellar medium. This can be expressed by the following identity, where the IMF is represented by $\varphi(m)$:
\begin{equation}
\frac{d^{3}N}{dm\,dV\,dt} = \dot{\rho}_{\star}(t)\cdot \varphi(m),
\label{eq:imf}
\end{equation}
where $dm$, $dV$, and $dt$ represent the infinitesimal elements of mass, volume, and time, respectively. 

The IMF is defined as the mass distribution function at the time of star formation, quantifying the fraction of stars within the mass interval $m, m+dm$. This function is normalized and classically described by the following equation \citep{Hennebelle_2024, Schneider_2016, Matteucci_2012, Kroupa_2002}:
\begin{equation}
\varphi(m)=\frac{m^{-(1+x)}}{\int_{m_{inf}}^{m_{sup}}m\cdot m^{-(1+x)}dm}.
\label{imf_normalizada}
\end{equation}

The IMF exponent $x$ in equation \ref{imf_normalizada} regulates the mass distribution towards larger- or smaller-mass concentrations. The exponent used in the ``classical'' IMF, the so-called  Salpeter IMF, is $x=1.35$ \citet{Salpeter_1959}. Varying $x < 1.35$ favors the formation of more massive stars and, in this case, the IMF is called a ``top-heavy''. Conversely, moving $x > 1.35$  favors the formation of less massive stars. 
In this work, we employ four IMF slopes: two top-heavy cases ($x=0.75, 0.95$), the standard Salpeter slope ($x=1.35$), and, for comparison, a slope favoring low-mass stars ($x=1.55$).
% Due to their initial chemical composition, Pop III stars tend to be more massive, which justifies the choice of a top-heavy IMF.

Mass is the dominant factor in determining a star's lifespan and can be estimated by the following relation:
\begin{equation}
    log\left(\tau_{m}\right) \approx 10-3.44,log\left(\frac{m_{\star}}{M_{\odot}}\right)+\left[log\left(\frac{m_{\star}}{M_{\odot}}\right)\right]^{2}.
    \label{tau_m2}
\end{equation}
This expression agrees with other estimates \citep{Rauscher_2020, Raiteri_1996} and has been used by other authors 
\cite{Gribel_2017, Pereira_Miranda_2011, Pereira_Miranda_2010, Copi_1997, Scalo_1986}. 
The nucleosynthetic contribution of a star of mass $m$ is intrinsically linked to its lifetime, $\tau_{m}$, during which the star processes and subsequently ejects enriched material into the interstellar medium.

Stellar remnants (white dwarfs, neutron stars and black holes) are created at the end of a stellar evolution process and usually store part of the chemically enriched material produced throughout the evolution of the star. In this sense, they consume enriched material and, in a different fashion of the cosmic BAR (but with the same consequences), contribute to the attenuation of metallicity.
Stellar remnant masses used in this work are determined following the prescriptions of \citet{Spera_2015} and  \citet{Pereira_Miranda_2010}:
\begin{enumerate}
    \item stars with masses $m < 1M_{\odot}$ have lifetimes equal to or greater than the Hubble time and therefore do not leave stellar remnants at this moment of the cosmic evolution; 
    \item Pop III stars with masses $\sim 140-260 M_{\odot}$ would evolve into pair-instability supernovae (PISN) \citep{Welsh_2024_a_survey, klessen_review_2023, Heger_Woosley_2010} and,
    \item stars with masses greater than $260 M_{\odot}$, in which the pair production process in PISN is insufficient to disassemble the star, ultimately collapses into a black hole without contributing to chemical enrichment \citep{Heger_Woosley_2010, Heger_Woosley_2002}. 
\end{enumerate}

IMBHs act as mass sinks and, in this work, their mathematical implementation is incorporated directly into the stellar yield function $P_X(m)$, for any element $X$ and remnant mass $m_r$. For stars with masses in the range $260 M_{\odot} < m \leq 10^4 M_{\odot}$, we define $P_X(m) = 0$ and the progenitor star is treated as the stellar remnant itself $m_r = m$. Physically, this implies that the entire stellar mass is locked within the resulting black hole, preventing its return to the interstellar medium and removing it from the chemical enrichment cycle, although without \textbf{explicitly} quantifying the cosmic accretion rate for black holes.

Recently, \citet{Miranda_2025} presented a possible solution to the cosmological lithium problem by demonstrating agreement between Standard Big Bang Nucleosynthesis predictions and the abundances observed in stars of the so-called Spite plateau. In particular, \citet{Miranda_2025} adopted the chemical model developed by \citet{Corazza_2022} and considered $260\,M_{\odot}$ as the upper mass limit for stellar populations capable of contributing to the chemical enrichment of the Universe. This same upper limit is used in the present work for stars that contribute to the nucleosynthesis of elements. Specifically, the stellar yield function $P_X(m)$ is considered non-zero only for masses up to $260\,M_{\odot}$. Beyond this limit, objects are treated as black holes, not returning any processed material to the interstellar medium.

%Recently, \citet{Miranda_2025} also adopted $260\,M_{\odot}$ as the upper mass limit for stellar populations contributing to the enrichment of the interstellar medium within the context of cosmic chemical evolution. In this work, stellar objects with masses up to $10^{4}\,M_{\odot}$ are assumed to be black holes acting as matter sinks, whose presence at high redshifts is predicted by other works \citep{Sanati_2025, Pereira_Miranda_2011}.

%\alex{Porque essa particularidade das Pop III com $M >260 M_{\odot}$? Sem rotação? Acho que o comentário tem que discutir também todas as pop III com massas capazes de evoluir no intervalo $20 > z > 6$ e falar que elas, com massas maiores que 8-10 $M_{\odot}$, vao contribuir durante muitas gerações até o final da sua formação em z=6. Então vc particulariza para esses 2 casos: PISN e M maior que 260 sem rotação. }

%Se há rotação, massas maiores que 260 podem enriquecer o meio, elas não colapsariam como propõem Heger e Woosley, por causa da força centrífuga.}

% \section{Mathematical Formulation of the Model}
\section{Model description}
\label{sec:themodel}

The evolution of cosmic chemical abundances can be quantified by considering the amount of material that enters and exits dark matter halos. In this sense, the temporal variation of the gas mass density ($\dot \rho_{g}(t)$) in dark matter halos depends on the amount of primordial baryonic matter ($a_b(t)$) that is gravitationally attracted to the dark matter halos, the amount of mass that is removed for star formation ($\dot \rho_{\star}(t)$), and finally, the amount of mass that returns to the medium through mass ejection from supernovae or stellar winds. Mathematically, this can be expressed as:

\begin{equation}
\frac{d\rho_{g}(t)}{dt}=-\dot{\rho}_{\star}(t)+\frac{d^{2}m_{ej}(t)}{dV\,dt}+a_{b}(t),
\label{densidade_gas}
\end{equation}

\noindent The cosmic rate of ejected stellar material is given by \citep{Tinsley_1973}:
\begin{equation}
\frac{d^{2}m_{ej}(t)}{dV\,dt}=\int_{m(t)}^{m_{sup}}\left(m-m_{r}\right)\dot{\rho}_{\star}\left(t-\tau_{m}\right)\varphi(m)dm.
\label{massa_ejet}
\end{equation}
% Note that the upper limit of integration $m_{sup}$ establishes the largest mass value present in the model; in this case, it is assumed that $m_{sup} = 10^{4}M_{\odot}$. In contrast, the lower limit of integration is a function of cosmic time ($m(t)$), which determines the mass of the star whose lifetime is $t$. It is advisable in this case to take $m(t)$ as the inverse of the function $\tau_m$. Furthermore, note that Equation \ref{massa_ejet} represents a cosmic quantity; that is, it does not quantify the ejected rate by a single star but rather by a set of stars over the evolution of the universe. 
It is worth noting that equation \ref{massa_ejet} represents a cosmic quantity, quantifying the ejected rate of all stars at a given moment in the evolution of the universe. Substituting \ref{massa_ejet} into \ref{densidade_gas} and doing some algebra leads to the differential equation that governs the temporal evolution of the abundance of an element $X$ in the gas of dark matter halos already used in previous studies \citep{Macedo_2026, Miranda_2025, Corazza_2022}:

\begin{multline}
        \frac{d\rho_{X}(t)}{dt} =  \int_{m(t)}^{m_{sup}}\left\{ \left(m-m_{r}\right)Z_{X}\left(t-\tau_{m}\right)+P_{X}(m)\right\} \times \\
       \dot{\rho}_{\star}\left(t-\tau_{m}\right)\varphi(m)dm -Z_{X}(t)\dot{\rho}_{\star}(t),
       \label{abundancia}
\end{multline}

\noindent with the initial condition of zero metallicity at redshift $z=20.$ 
% \refa{An upper mass limit of $m_{sup}=10^{4}M_{\odot}$ is adopted. However, for stellar masses above $260M_{\odot}$, the stellar yield function is set to zero and the entire stellar mass is treated as a remnant, i.e. as an object that does not return mass to the interstellar medium, as previously described.} Note that the fraction of element $X$ present in the gas at time $t$ is equal to what was present at the time of the star's birth $(Z_X(t-\tau_m))$ plus what was produced by the star $(P_X(m))$, but diluted in the ejected material ($m-m_r$). 
As described in Section 2, we adopt $m_{sup}=10^{4}M_{\odot}$ as the mass upper limit, stressing that the  stellar yield function is set to zero for stellar masses above $260M_{\odot}$. In this situation, the entire stellar mass is treated as a remnant, with its entire mass not returning back to the interstellar medium. Note that the fraction of element $X$ present in the gas at time $t$ is equal to what was present at the time of the star's birth $(Z_X(t-\tau_m))$ plus what was produced by the star $(P_X(m))$, but diluted in the ejected material ($m-m_r$). 
It should be noted that while the CSFR results are displayed in terms of $\log_{10}(\dot\rho_{\star})$ (following Equation \ref{melhor ajuste}), the function $\dot\rho_{\star}(z)$ utilized in Equations \ref{densidade_gas}, \ref{massa_ejet}, and \ref{abundancia} corresponds to the physical rate, expressed in $M_{\odot}\,\text{yr}^{-1}\,\text{Mpc}^{-3}$.

The Pop III stellar yields adopted in this work are based on the classical prescriptions of \citet{Heger_Woosley_2002, Campbell_Lattanzio_2008_CL08, Heger_Woosley_2010}. It should be noted that the objective of this work is not to discuss the underlying stellar evolution models for primordial stars, but rather to employ them, together with other physical processes, as an (important) ingredient within a chemical enrichment framework.

Equation \ref{abundancia} is then solved numerically using the \texttt{odeint} routine in \texttt{Python}. %, and the results are consistent with the numerical solution obtained using a fifth-order Runge--Kutta method.
Finally, the abundance of elements is generally assessed on the scale [X/H], defined by:
\begin{equation}
    \left[\frac{X}{H}\right]=log\left[\frac{n(X)}{n(H)}\right]_{gas}-log\left[\frac{n(X)}{n(H)}\right]_{\odot},
\end{equation}
The following section presents the results, compared to the solar abundances from \citet{Asplund_2021, Asplund_2009}.
%--------------------------------------------------------------------
%--------------------------------------------------------------------
%--------------------------------------------------------------------
%--------------------------------------------------------------------
\section{Results and discussion}
\label{sec:results}

\begin{figure}[h!]
     \centering
     \begin{subfigure}[b]{0.495\textwidth}
          \centering
          \includegraphics[width=\textwidth]{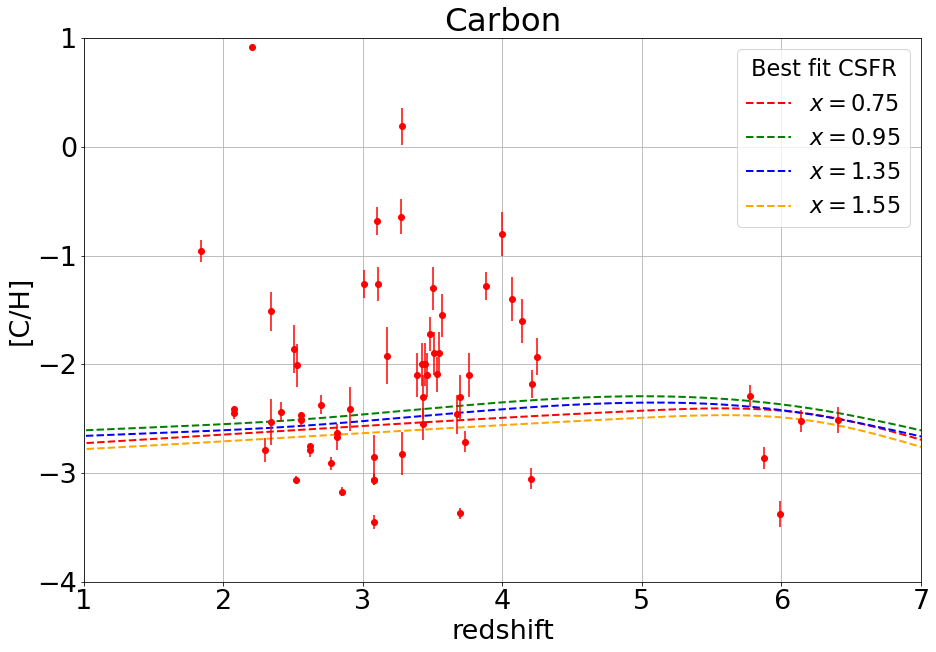}
     \end{subfigure}
     \hfill
     \begin{subfigure}[b]{0.495\textwidth}
          \centering
          \includegraphics[width=\textwidth]{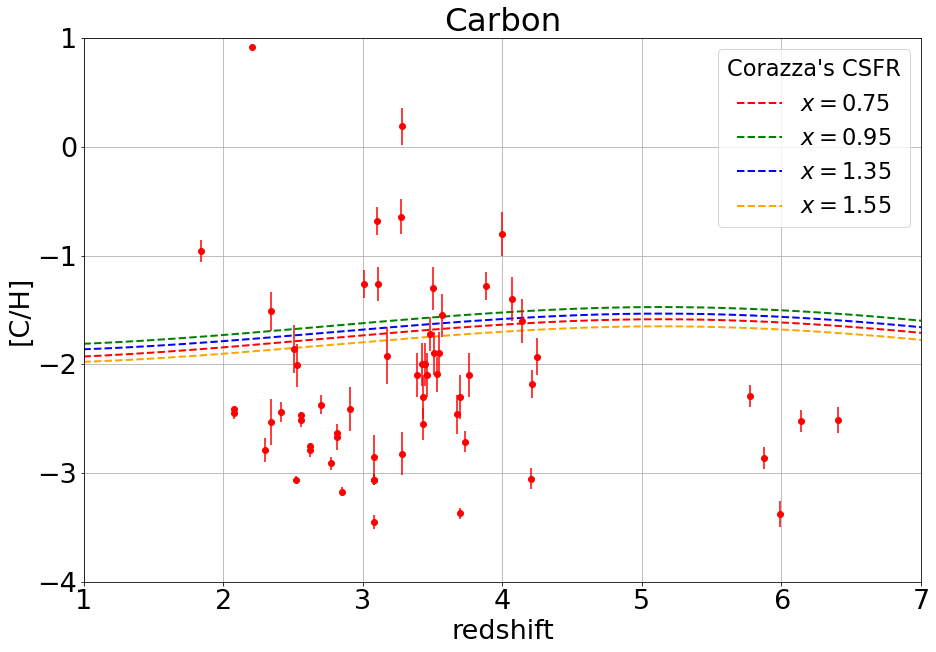}
     \end{subfigure}

     \begin{subfigure}[b]{0.495\textwidth}
          \centering
          \includegraphics[width=\textwidth]{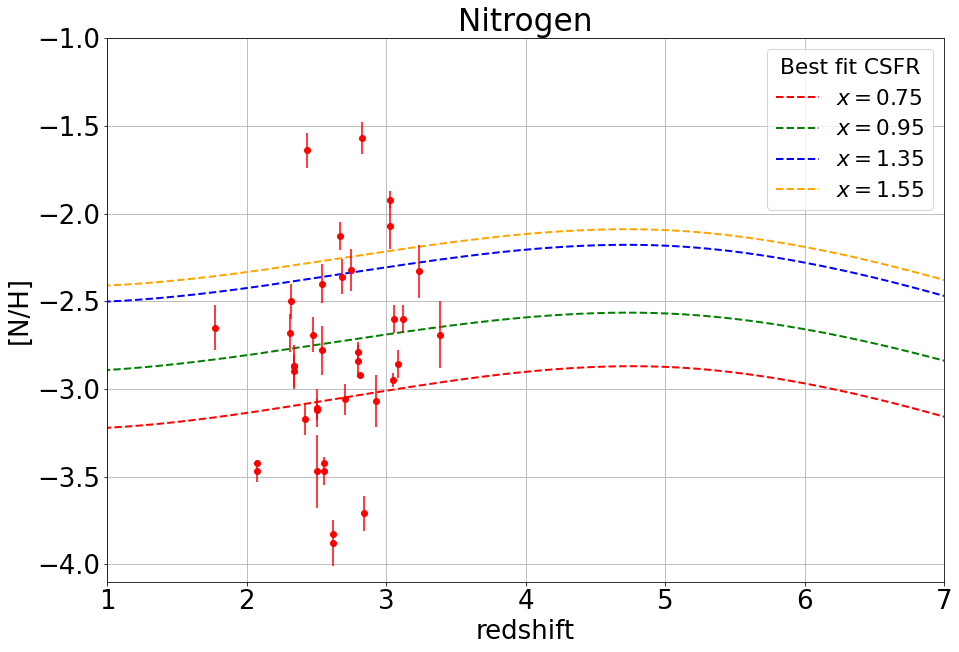}
     \end{subfigure}
     \hfill
     \begin{subfigure}[b]{0.495\textwidth}
          \centering
          \includegraphics[width=\textwidth]{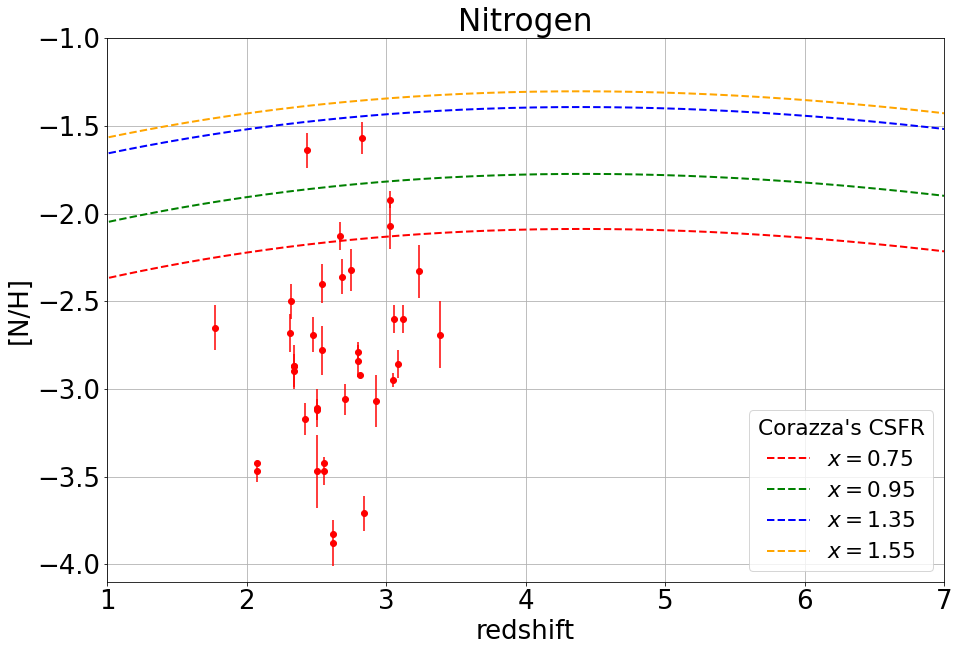}
     \end{subfigure}

     \begin{subfigure}[b]{0.495\textwidth}
          \centering
          \includegraphics[width=\textwidth]{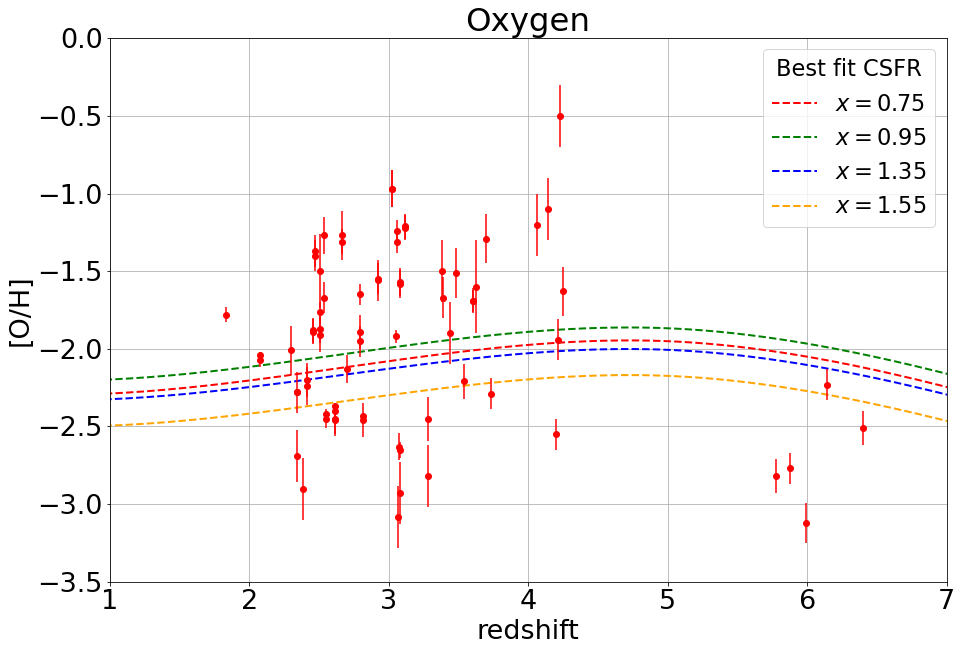}
     \end{subfigure}
     \hfill
     \begin{subfigure}[b]{0.495\textwidth}
          \centering
          \includegraphics[width=\textwidth]{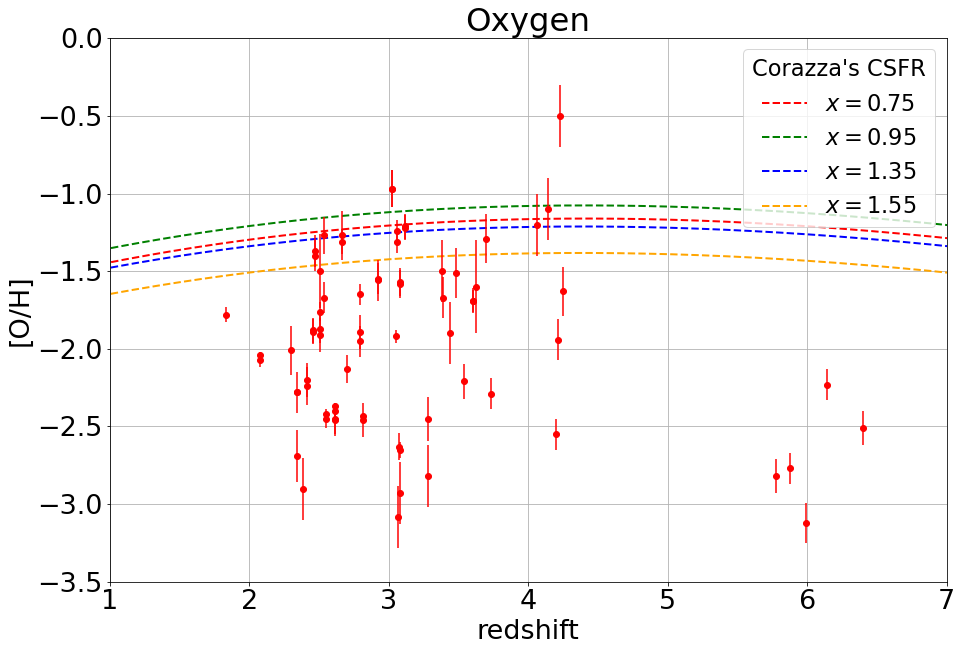}
     \end{subfigure}

     \caption{Chemical enrichment model predictions (dashed curves) for different values of $X/H$ compared to ASQS data (red points). Left panels show the Best-fit CSFR + IMBHs model, while right panels show Corazza's CSFR + IMBHs. Observational data compiled from \citep{Welsh_2024_a_survey, Sodini_A_2024, Saccardi_2023, Welsh_L_2022, Poudel_2021, De_Cia_2016, Dutta_2014, Zafar_2014, Kulkarni_2012, Cooke_2011_the_most_metal, Cooke_2011_b, Penprase_2010, Pettini_2008, Petitjean_2008, Dessauges_2003, Dessauges_Zavadsky_2002, Prochaska_1998}}
     \label{fig_resultados}
\end{figure}

The results are presented in two columns in Figure \ref{fig_resultados}: accounting for IMBHs in both scenarios, the left column shows the best-fit CSFR within the previously discussed model, while the right column employs the CSFR from \citet{Corazza_2022}. In all cases, we consider the four IMF exponents as described in the legend.
According to the $\Lambda$CDM model, at $z \sim 6$ the Universe was about $\sim 940$ Myr. %after the Big Bang. 
At this epoch, the presence of C, O is detected in ASQS, suggesting that the nucleosynthetic source responsible for these elements operates on very short timescales, compared to the age of the Universe, starting at $z=20~(\sim 180$ Myr) years. This scenario is consistent with the presence of Pop III stars, which efficiently produce C, N, and O, according to the stellar yield models adopted \citep{Heger_Woosley_2010, Campvell_Lattanzio_2008_CL08, Heger_Woosley_2002}.

Pop III stars cease their activity around $z \sim 6$. At this point, the curves for all elements (C, N, O) show a decline and the gas is continually diluted by the CBAR, which accretes primordial H and He into the gas. Furthermore, due to cosmic expansion, the BAR also decreases after $z \sim 11$ (Equation \ref{BAR_equation}). This allows for a slight enhancement in CNO abundances, particularly for N in the first column. The N evolution, in particular, is driven by the delayed contribution from lower-mass stars. %By favoring the presence of more massive IMBHs with lower $x$ values, chemical elements also tend to undergo higher mass sequestration. 
The choice of lower values for $x$ favors the presence of more massive IMBHs with, on its turn, increases the mass sequestration of chemical elements from the cosmic environment. This establishes a competition between stellar yields, CBAR-driven dilution, and the removal of gas via IMBH sink terms.

The yields of C and O from Pop III stars, originating from either stellar nucleosynthesis or explosive supernova events, are considerably higher than those of N, with O being the most abundant element produced. Consequently, it can be inferred that supernovae are more efficient at dispersing N into the interstellar medium than at synthesizing it \citep{Kobayashi_2020, Nomoto_2013, Pettini_2008, Heger_Woosley_2010, Heger_Woosley_2002}.

Within the Pop III star framework, O is an element of rapid production via massive stars. This leads to $N/O$ and $C/O$ ratios increasing alongside gas metallicity as further enrichment occurs, primarily due to the contribution of Asymptotic Giant Branch (AGB) stars to these elements. Specifically, intermediate-mass AGB stars ($\sim 4 - 8 M_{\odot}$) undergoing the CNO nuclear cycle are estimated to be responsible for $74\%$ of the N and $49\%$ of the C observed in absorbers \cite{Handbook_nuclear_physics_2023_88, Kobayashi_2020, Nomoto_2013, Pettini_2002}. Assuming Pop III star formation begins at $z=20$ (approximately $180\,Myr$ after the Big Bang), stars in the $\sim4-8\,M_{\odot}$ range would only contribute to the chemical enrichment after the evolution of the more massive stars. This time delay could explain the deficit of N observed in high-redshift ($z \sim 6$) absorption systems compared to O and C \citep{Rafelski_2014,Petitjean_2008, Rafelski_2012}. Furthermore, there is a lack of observational data for N at $z \gtrsim 3.5$ in these systems, partly due to the practical difficulties in measuring this element caused by line blending within the Lyman-alpha forest \citep{Sodini_A_2024, Pettini_2002, Calura_2003}.

In environments located at $z \sim 2$--$4$ ($\sim 2.5$--$3.5$ billion years), when Pop III star formation is assumed to have already ceased, the model provides a reasonable fit for the elements C, N, and O, correcting the excess reported by \citet{Corazza_2022}. Within this redshift interval, it is worth emphasizing that the model reproduces reasonably well the most metal-poor ASQS, which may indicate %a nucleosynthetic contribution 
some apport from more massive Pop II stars that evolve on shorter timescales and its consistent to \citet{Welsh_2024_a_survey, Sodini_A_2024} which do not discard the role of Pop II in the same context. Regarding the observations of C and O at $z \sim 6$, the model proposed in this work shifts the enrichment curve downward, showing strong consistency with observational data. Although a slight excess persists for O at $z \sim 6$, the model provides a highly accurate fit for C, and this small excess does not undermine the model's validity, but instead highlights the localized complexities involved in real-world chemical enrichment. It is reasonable to infer that the excess reported by \citet{Corazza_2022} stems from their high CSFR (red curve in Figure \ref{fig_csfr}), which may not be suitable for absorbers, as they typically exhibit lower star formation rates than galaxies.

%\alex{Entendi a ideia mas acho que precisa escrever melhor, de maneira mais clara e com mais informação e contexto. Se possíve, com mais referêwncias. Usar só o Isobe dá a ideia de um viés.}
%In agreement with our results for C and O in ASQS, \citet{Welsh_2024_a_survey} find that abundances are consistent with a Salpeter IMF, while noting a preference for a more top-heavy slope. 
Other results in  the literature (see, e.g. \cite{Welsh_2024_a_survey}) find that abundances are consistent with a Salpeter IMF, while noting a preference for a more top-heavy slope, which directly suppports our results for C and O in ASQS. Although these represent distinct chemical enrichment models and different observational data, they exhibit consistent behavior regarding the IMF exponent. This supports our approach of utilizing a modified IMF to investigate the early chemical enrichment of the interstellar medium, particularly when considering mass sequestration by IMBHs. The top-heavy IMF in our model promotes IMBH formation, leading to higher mass consumption and a reduction in the previously mentioned excess which motivated this study.

The upper mass limit was set to encompass a broad range of astrophysical objects. The IMF is integrated up to $10^4 M_{\odot}$ to account for the masses of either stellar bodies or IMBHs, ensuring that the model captures a broad spectrum of massive sinks, from the massive Pop III stars to the resulting IMBHs. %The sink effect is thus a direct consequence of the IMF weighting: while these stars are born (consuming gas), they die with no mass ejection (null yields), which naturally leads to the attenuation of the chemical enrichment.
The capture of matter by IMBHs (the ``sinking'') is thus a direct consequence of the IMF weighting: while these stars are born (consuming gas), they die with no mass ejection (null yields), which naturally leads to the attenuation of the chemical enrichment. O is the most abundant species among the CNO elements, as it is synthesized in large quantities by massive stars. Consequently, the total mass sequestered by IMBHs is expected to be higher for this element. Our results for O, particularly when adopting $x=0.95$ and $x=0.75$, demonstrate that increasing the proportion of massive stars (lower $x$) leads to a higher mass consumption due to the increased frequency of IMBHs. C, the second most abundant element, is produced in smaller amounts by high-mass stars, which results in a smaller scattering across the models. Regarding N, its production is dominated by lower-mass stars; this is reflected in the data, where attenuation becomes more evident when IMBHs are included in scenarios with lower $x$ values, which favor the presence of more massive sinks.

The presence of black holes acting as metallicity attenuators in galactic environments has already been suggested by \citet{Isobe_2023}. In this context, the present work reinforces the same hypothesis, but extended to other types of astrophysical environments. Stellar formation and, consequently, chemical enrichment, is also affected by radiative feedback from massive black holes \citep{Zhu_2025}. However, due to the inherent complexity of integrating such processes into chemical evolution models, we have not included radiative effects in this work. Instead, we treat IMBHs exclusively as matter sinks. The results obtained in this work are summarized in Table \ref{tab_sintese}.

\begin{table}[h!]
\centering
\small
\begin{tabularx}{\textwidth}{@{} l X X @{}}
\toprule
\textbf{Element} & \textbf{Corazza's CSFR + IMBHs} & \textbf{Best-fit CSFR + IMBHs} \\ \midrule
Carbon & Tracks around the median upwards; large excess at $z \sim 6$. & Intercepts around the data median at $z \sim 3$; consistent with $z = 6$ data. \\ \midrule
Nitrogen & Fails to predict the low-metallicity regime ($-3.5 < [N/H] < -2.5$). & Reduction of metal excess; the model successfully describes the main ASQS. \\ \midrule
%Oxygen & Curves act as an upper limit with an excess at $z \sim 6$. & Curves track the evolutionary trend of the observations, with better agreement at $z \sim 6$. \\ \bottomrule
Oxygen & Curves act as an upper limit with an excess at $z \sim 6$. & Curves track  evolution of observations: better agreement at $z \sim 6$. \\ \bottomrule
\end{tabularx}
\caption{Comparison of models for C, N, and O.}
\label{tab_sintese}
\end{table}

\section{Conclusions}
\label{sec:conclusions}

Throughout this work, we presented a semi-analytical model of cosmic chemical enrichment that incorporates Pop III stars as a key source of early metal nucleosynthesis, together with baryonic accretion from primordial nucleosynthesis and the presence of IMBHs, both of which act as metal attenuators. It is worth noting that the presence of IMBHs in high-redshift is supported by other studies \citep{Pereira_Miranda_2011, Pereira_Miranda_2010}.

While the model of \citet{Corazza_2022} seems to act as an upper limit by overestimating metal abundances across a range of redshifts, the inclusion of IMBHs in our model effectively sequesters a significant fraction of these metals. Our results demonstrate that IMBHs exert a substantial impact on cosmic chemical evolution that should not be overlooked. This process mitigates the aforementioned excess and brings the curves into alignment with the region of highest statistical density in the ASQS data. Furthermore, the results obtained here are consistent with previous studies indicating that a subset of ASQS preserves chemical signatures associated with Pop III stars \citep{Salvadori_2023, Saccardi_2023, Vanni_2023, Kobayashi_2011}. At lower redshifts ($z \lesssim 4$), however, the contribution of Pop II stars becomes increasingly important, especially for elements requiring more advanced nucleosynthetic channels \citep{Sodini_A_2024, Welsh_L_2022}. Due to the implementation complexity arising from their existence across different metallicity classes, Pop II stars were not implemented in this paper.

In light of recent JWST observations, the existence of massive Pop III stars has been increasingly discussed as a plausible explanation for several early-Universe phenomena \citep{Nandal_2024_3000M, Nandal_2024, Maiolino_2024, Harikane_2023, Chantavat_2023, Woods_2021, Liu_2020} while a growing body of literature investigates extreme supermassive primordial stars reaching tens of thousands of solar masses \citep{Nandal_2026, Nandal_2025_GS3073, Chantavat_2023, Woods_2021}, which can be interpreted as potential birth sites for IMBHs. The present study reinforces this interpretation by demonstrating that scenarios featuring high-mass Pop III stars provide a compelling and consistent explanation for the chemical abundance patterns observed in ASQS. While alternative pathways cannot be fully excluded, this work highlights that early chemical enrichment is driven by a complex interplay between the rapid yields of Pop III stars and the delayed contributions of AGB stars, modulated by both primordial gas accretion and mass sequestration by IMBHs.

%The formation of such massive primordial stars can be understood in terms of the physical properties of primordial gas. The lack of efficient metal-line cooling implies higher characteristic fragmentation masses, favoring the formation of massive stars in order to overcome thermal pressure during gravitational collapse. In this context, the presence of intermediate-mass or very massive black holes at high redshifts further supports a picture in which early structure formation is dominated by high-mass objects \citep{Ziparo_2024, Larson_2023, Farina_2022}.

We demonstrate that IMBHs play a significant role as metallicity attenuators, where mass sequestration by these objects competes directly with stellar nucleosynthesis and CBAR-driven dilution. 
Our results indicate that mass sequestration by IMBHs is a determinant factor in reconciling theoretical predictions with observational data. This mass-locking effect effectively mitigates the CNO overproduction by preventing the return of processed material to the chemical enrichment cycle. Consequently,
while the current model is consistent with the observed abundances of C, N and O for the range of redshift investigated (with some O excess at $z\sim6$) the precise impact of IMBHs, particularly concerning radiative feedback, accretion rate and more diverse metallicity classes of stars remains a key area to be investigated in future studies.

On the other hand, the eROSITA Final Equatorial Depth Survey (eFEDS) has identified a black hole consuming mass at a rate exceeding the predicted theoretical limit, which may challenge accretion models in the context of cosmic chemical enrichment \cite{Obuchi_2026}. Regarding the cosmic black hole accretion rate, the works of \citet{Luberto_2025} and  \citet{Pereira_Miranda_2011} may be useful for this advancement. %Therefore, this article reinforces the necessity of considering black hole sink terms as a fundamental component in the chemical evolution of the Universe.  
As a final note, this article stresses the necessity of considering black hole sink terms as a fundamental component in the chemical evolution of the Universe.  

\vspace{0.5cm}
\noindent \textbf{Acknowledgements}: The authors acknowledge the Brazilian Ministry of Science, Technology and Innovation (MCTI)
and the Brazilian Space Agency (AEB) who supported the present work under the PO 20VB.0009. CAW explicitly acknowledges the CNPq grant 407446/2021-4.

\vspace{0.5cm}
\noindent \textbf{Declaration}: The authors declare that they did not receive any specific funding to produce this work. 

\bibliography{references}% common bib file

\end{document}